                        \newif\ifboyscout                         %%
                        \boyscoutfalse %% commented, WWW/boyscouts %%
                        \newif\ifpreparepdf                       %%
                        \preparepdftrue % hyperlinked pdf default %%
                        \newif\ifhighlightedits
                        \highlighteditsfalse

\documentclass{jfm}

\ifhighlightedits
    
\else
    
\fi    
    
        \ifboyscout % if draft, display comments in text
   \newcommand{\PCedit}[1]{{\color{blue}#1}}
   \newcommand{\PC}[2]{\begin{quote}\PCedit{[#1 Predrag] #2}\end{quote}}
   \newcommand{\RLDedit}[1]{{\color{red}#1}}
   \newcommand{\RLD}[2]{\begin{quote}\RLDedit{[#1 Ruslan] #2}\end{quote}}
   \newcommand{\BBedit}[1]{{\color{green}#1}}
   \newcommand{\BB}[2]{\begin{quote}\BBedit{[#1 Burak] #2}\end{quote}}
   \newcommand{\ESedit}[1]{{\color{magenta}#1}}
   \newcommand{\ES}[2]{\begin{quote}\ESedit{[#1 Evangelos] #2}\end{quote}}
   \newcommand{\HC}[2]{\begin{quote}[#1 Hugues] #2\end{quote}}

        \else  % drop comments
   \newcommand{\PC}[2]{}{}
   \newcommand{\PCedit}[1]{#1}
   \newcommand{\RLD}[2]{}{}
   \newcommand{\RLDedit}[1]{#1}
   \newcommand{\BB}[2]{}{}
   \newcommand{\BBedit}[1]{#1}
   \newcommand{\ES}[2]{}{}
   \newcommand{\ESedit}[1]{#1}
   \newcommand{\HC}[2]{}{}
        \fi

\newcommand{\Lg}{\ensuremath{T}}   % 2014-04-04 prettier Lie algebra generator
\newcommand{\LieEl}{\ensuremath{g}}  %  a group element (often replaced by \matrixRep)
\newcommand{\rf}     [1] {~\cite{#1}}
\newcommand{\rfp}[1] {~\citep{#1}}

\newcommand{\refeq}  [1] {(\ref{#1})}
\newcommand{\reffig} [1] {fig.~\ref{#1}}
\newcommand{\refFig} [1] {Fig.~\ref{#1}}

\newcommand{\beq}{\begin{equation}}
\newcommand{\continue}{\nonumber \\ }

\newcommand{\eeq}{\end{equation}}
\newcommand{\ee}[1] {\label{#1} \end{equation}}
\newcommand{\bea}{\begin{eqnarray}}

\newcommand{\eea}{\end{eqnarray}}

\newcommand{\ie}{{i.e.}}            % APS

\newcommand{\etal}{{\em et al.}}    % etal in italics, APS too

\newcommand{\statesp}{state space}
\newcommand{\Statesp}{State space}
\newcommand{\zeit}{\ensuremath{\tau}}  %time variable
\newcommand{\po}{periodic orbit}

\newcommand{\rpo}{relative periodic orbit}
\newcommand{\bseq}{\begin{subequations}}
\newcommand{\eseq}{\end{subequations}}

\newcommand{\NSe}{Navier-Stokes equations}
\newcommand{\Reynolds}{\textit{Re}}  % Reynolds number
\newcommand{\stateDsp}{state-space}

\newcommand{\mslices}{method of slices}

\newcommand{\sspRed}{\ensuremath{\hat{\ssp}}}    % reduced state space point Jan 2012
\newcommand{\velRed}{\ensuremath{\hat{\vel}}}    % ES reduced state space velocity Jan 2012
\newcommand\flowRed[2]{{\hat{f}^{#1}(#2)}}

\newcommand{\slicep}{{\ensuremath{\sspRed'}}}   % slice-fixing point Jan 2012
\newcommand{\sliceTan}[1]{\ensuremath{t'_{#1}}}    % group orbit tangent at slice-fixing
\newcommand{\groupTan}{\ensuremath{t}}    % group orbit tangent
\newcommand{\Group}{\ensuremath{G}}         % Predrag Lie or discrete group

\newcommand{\gSpace}{\ensuremath{{\phi}}}   % MA group rotation parameters
\newcommand{\gSpaceRed}{\ensuremath{{\hat{\phi}}}}   % MA group rotation parameters
\newcommand{\KS}{Kuramoto-Siva\-shin\-sky}

\newcommand{\eigExp}[1][]{
     \ifthenelse{\equal{#1}{}}{\ensuremath{\lambda}}{\ensuremath{\lambda^{(#1)}}}}
\newcommand{\eigRe}[1][]{
     \ifthenelse{\equal{#1}{}}{\ensuremath{\mu}}{\ensuremath{\mu^{(#1)}}}}
\newcommand{\eigIm}[1][]{
     \ifthenelse{\equal{#1}{}}{\ensuremath{\omega}}{\ensuremath{\omega^{(#1)}}}}

\newcommand\flow[2]{{f^{#1}(#2)}}
\newcommand{\vel}{\ensuremath{v}}   % state space velocity
\newcommand{\ssp}{\ensuremath{a}}                % state space point
\newcommand{\fFslice}{first Fourier mode slice}

\newcommand{\inprod}[2]{\left\langle #1 ,\, #2 \right\rangle}
\newcommand{\PoincS}{\ensuremath{{\cal P}}}  % symbol for Poincare section
\newcommand\flowP[2]{{\hat{f}_{\PoincS}^{#1}(#2)}}
\newcommand{\ProjPsect}{P}

            \newif\ifdraft \newif\ifcolorfigs \newif\ifJFM
            \draftfalse \colorfigstrue \JFMtrue

\ifJFM
  \usepackage{upmath}
\fi

\ifboyscout
\fi

\usepackage{ifpdf}
\ifpdf
  \usepackage[pdftex]{color,graphicx}
  \usepackage{epstopdf}
\else
  \usepackage[dvips]{color,graphicx}
\fi

\usepackage{overpic}
\usepackage{float}
\usepackage{natbib}
\usepackage{amsmath,amsfonts,amssymb,amsbsy,amscd,amsgen}
\usepackage{gensymb}
\usepackage{url}
\usepackage{subfigure}
\usepackage{ifthen}
\usepackage{appendix}
\graphicspath{{figs/}}  %% directories with graphics files

\usepackage[active]{srcltx}

\ifpreparepdf % hyperlinked pdf, keep homepage flexible:

\else  %% prepare for postscript printing:

\fi

\ifcolorfigs
  \usepackage[pdftex,colorlinks]{hyperref} %% hyperlinks, LAST package called
\else
\fi

\title[Heteroclinic path to spatially localized chaos]
{
Heteroclinic path to spatially localized chaos in pipe flow
}
\author[
N.\ B.\ Budanur %,
and
B.\ Hof
        ]
{
N.\ns B.\ns B\ls U\ls D\ls A\ls N\ls U\ls R$^{1,2}$%,\ns
\ns
\and
B.\ns H\ls O\ls F$^{1}$
}
\affiliation{$^1$ IST Austria,
                  3400 Klosterneuburg Austria\\[\affilskip]
             $^2$ Kavli Institute for Theoretical Physics,
                  UC Santa Barbara, Santa Barbara, CA 93106}

\pubyear{2017}
\begin{document}
\maketitle

\date{\today}

        \abstract{
        In shear flows at transitional Reynolds numbers, localized 
        patches of turbulence, known as puffs, coexist with the laminar 
        flow. Recently, Avila \etal , Phys. Rev. Let. \textbf{110}, 
        224502 (2013) discovered two spatially localized relative 
        periodic solutions for pipe flow, which 
        appeared in a saddle-node bifurcation at low speeds.
        Combining slicing methods for continuous symmetry reduction
        with Poincar\'e sections for the first time in a shear flow 
        setting, we compute and visualize the unstable manifold of 
        the lower-branch solution and show that it contains a 
        heteroclinic connection to the upper branch solution. 
         Surprisingly this connection even persists far above the bifurcation 
         point and appears to mediate puff generation, providing a dynamical 
         understanding of this phenomenon.
        }

\section{Introduction}

In pipe flow, turbulence first appears in localized regions known as puffs.
Puffs propagate downstream and
eventually decay back to the laminar state or split to give birth 
to a new puff. Numerical and laboratory experiments in pipe 
flow\rfp{AMdABH11} have shown that transition to sustained turbulence in 
a circular pipe happens when the rate of puff splitting exceeds that 
of decaying. Further research\rfp{LSAJAH16} has established 
that the dynamics and interplay of such localized turbulent domains
give rise to a non-equilibrium phase transition. 
Instead of this stochastic point of view, we here apply the
complementary deterministic dynamical systems approach in order to 
unravel details of puff formation.

The time evolution of a fluid flow can be thought of as a trajectory
in an infinite dimensional space.
In this \textit{\statesp}, a state of the fluid is a point, and its motion 
is a one-dimensional curve. While \rf{hopf48} articulated this 
dynamical viewpoint of turbulence, it has only
recently been computationally feasible to tackle turbulence 
from this perspective. From this geometrical viewpoint, 
\statesp\ of a transitional shear flow contains a linearly stable
\textit{equilibrium}, laminar flow, and a 
\textit{chaotic saddle}, turbulence, 
which for the parameter regime considered here is of transient 
nature.

\Statesp\ geometry of a chaotic system is shaped by setwise 
time-invariant solutions, such as equilibria and periodic orbits; 
and their stable/unstable manifolds\rfp{DasBuch}. 
All of these solutions are unstable; thus the chaotic flow
transiently approaches an invariant solution 
following its stable manifold
and leaves its neighbourhood on its unstable manifold. 
This intuition suggests the search for invariant 
solutions as the first step of the turbulence problem from 
Hopf's perspective.

Numerically finding invariant solutions of the \NSe\ for canonical 
(Couette and Pouseuille) shear flow geometries
is a non-trivial task and for Newton-based methods
it is not obvious how to find a \textit{good} initial guess. Early 
studies\rfp{N90,W98,FE03} relied on the \textit{homotopy} method, 
in which a term is added to the \NSe\ that continuously maps a canonical flow
to some other flow where an invariant solution is known to exist;
subsequently this term is adiabatically turned off
while numerically continuing the invariant solution. 
A general feature of the invariant solutions found this way was
that they appeared as 
pairs in saddle node bifurcations at low Reynolds numbers 
(\Reynolds) or through further bifurcations of such saddle-node
pairs. Moreover, the lower-energy ones of these solutions appeared 
to belong to a \statesp\ region between laminar and 
turbulent dynamics. Studies\rfp{TI03,SYE05,SchEckYor07}, which 
focused on the \textit{edge of chaos} in shear flows, 
suggested that this separatrix between laminar and turbulent
dynamics as the stable manifold of an invariant set of solutions.
While in most cases, these solutions themselves 
exhibit complicated - albeit simpler than turbulence - dynamics, 
in some settings, they were particularly
simple such as an equilibrium or a time-periodic solution.
These developments 
suggested 
the laminar turbulent boundary as
a new starting point for invariant solution searches, 
which proved successful for 
pipe\rfp{duguet07}, plane Couette\rfp{SGLDE08}, and plane 
Poiseuille\rfp{ZamEck15} flows. 

Most of the studies we cited above were restricted to small 
computational domains called \textit{minimal flow units}\rfp{JM91} 
that are only large enough to capture essential statistics of turbulence. 
However, such small domains cannot capture streamwise localization
of turbulent spots, which are relevant to the onset of turbulence.
\rf{AvMeRoHo13} numerically studied the laminar-turbulent 
boundary in a 40-diameter-long periodic pipe, a domain large enough to 
exhibit localization. They discovered that when the flow is restricted 
to the invariant subspace of solutions that 
have 2-fold rotational and reflectional symmetries in azimuth,
the edge of chaos
is a single \textit{relative} periodic orbit. By numerical 
continuation 
they showed that this solution is the lower-branch of a saddle-node 
pair; akin to invariant solutions found in small domains.

	For a complete understanding of the \statesp\ geometry, numerical
	identification of dynamically relevant invariant solutions must be 
	followed by preparation of a \textit{catalogue} of possible motions 
	in their neighbourhood. This is a technically challenging task since
	the unstable manifolds of these solutions can be very
	high dimensional. Moreover, presence of continuous symmetries further
	complicates this by adding extra dimensionality. 
	The main technical contribution of the current work is to resolve this
	issue by combining the \mslices\ with Poincar\'e 
	sections for computation and visualization of the unstable 
	manifold of a \rpo . 
	Applying this technique we show for the aforementioned case of the localized 
    \rpo\ that a heteroclinic connection to the upper branch governs transition 
    and the formation of puffs. 
	The paper is organized as follows: 
In the next section, we describe the numerical procedure and 
introduce our notation. 
Technical aspects, the simple symmetry reduction 
scheme and the computation of unstable manifolds on a Poincar\'e 
section are presented in sections \ref{s-sym} and \ref{s-man}. 
The results are discussed in section \ref{s-con}. The 
main text is supplemented by Appendix \ref{s-proj}, where we derive 
projection operators, which we use in our computations.

\section{Numerical setup and notation}
\label{s-num}
For numerical simulations, 
we use the primitive-variable version of 
\texttt{openpipeflow.org} \rfp{WillKer09,openpipeflow}, 
which integrates the Navier-Stokes equations 
for fluctuations ${\bf u}$ around the
base  (Hagen-Poiseuille) solution. 
      The axial pressure gradient is adjusted throughout the
      simulation in order to ensure a constant flux equal
      to that of the base flow at a given 
      $\Reynolds = U D / \nu$,
      where $U$ is the mean axial velocity, $D$ is the
      pipe diameter, and $\nu$ is the kinematic viscosity 
      of the fluid.
The flow field satisfies 
the incompressibility condition 
$\nabla \cdot {\bf u} = 0$ throughout the volume, 
periodic boundary conditions 
${\bf u}(z, \theta, r) = {\bf u}(z + L, \theta, r)$ and
${\bf u}(z, \theta, r) = {\bf u}(z, \theta + 2 \pi, r)$
in axial and azimuthal directions,
and no-slip boundary condition
${\bf u}(z, \theta, r = D/2) = 0$
on pipe wall. 
In the numerical work presented here, the computational 
domain is $L=25D$-long and
we use $192$ and $16$ Fourier modes respectively in 
axial and azimuthal 
directions and $64$ finite difference points in the radial 
direction. 
\Reynolds\ is set to 1700 at which puffs have long 
(up to 1000 D/U) life times.

Pipe flow is \textit{equivariant} under streamwise translations 
\(\LieEl_z (l){\bf u}(z, r, \theta) = {\bf u}(z-l, r, \theta)\,,\)
azimuthal rotations
\(\LieEl_\theta (\phi){\bf u}(z, r, \theta) = {\bf u}(z, r, \theta-\phi)\,,\)
and the azimuthal reflection 
\(\sigma [u, v, w]\) \((z, r, \theta)= [u, v, -w](z, r, -\theta)\,,\)
where $u, v, w$ are velocity field components in axial, radial,
and azimuthal directions respectively. Following \rf{AvMeRoHo13}, we
restrict our study to the velocity fields
that are symmetric under rotation
by $\pi$, $\LieEl (0, \pi) {\bf u} = {\bf u}$,
and the reflection
$\sigma {\bf u} = {\bf u}$. Imposing reflection invariance
breaks the continuous rotation symmetry of the system, allowing
only for half-domain rotations. Therefore, the symmetry group of 
the system becomes 
$\Group = \{\LieEl_z(l), \LieEl_\theta(\pi/2)\}$. Note that rotation
by $\pi / 2$ is a rotation by $\pi$ in the subspace of azimuthally
doubly periodic fields.

For clarity,
we are going to use \statesp\ notation: 
Let, $\ssp (t)$ be a vector, containing
all $(3\times64\times192\times16)$
numerical degrees of freedom of the flow fields $[u,v,w]$. 
Evolution under \NSe\ implies a finite-time
flow mapping $\flow{\zeit}{\ssp (0)} = \ssp (\zeit)$ that takes a 
solution at time $0$ to a new one at time $\zeit$. In this formalism, 
equivariance under $\LieEl\!\in\!\Group$ means that the flow and the 
symmetry operation commutes, \ie\ 
$\LieEl \flow{\zeit}{\ssp (0)} = \flow{\zeit}{\LieEl \ssp (0)}$.
Throughout this article, our notation will not distinguish between 
an abstract group element and its particular representation. 
Thus, whenever a group action 
is present, its appropriate representation on the corresponding 
velocity fields is implied. 
Finally, we need to introduce an inner
product to use in our calculations. Our choice is
the standard ``energy norm'': Let $\ssp$ and $\ssp'$ be \statesp\ 
vectors corresponding to velocity fields ${\bf u}$ and ${\bf u'}$, 
we define the $L_2$ inner-product as 
$\inprod{\ssp}{\ssp'} = (1/2) \int {\bf u} \cdot {\bf u'}dV $; hence
$||\ssp||^2 = \inprod{\ssp}{\ssp}$ is the kinetic energy of $\ssp$.

A \rpo\ is a recurrence (after the orbit's period $T'_p$) up to 
a symmetry operation
\beq
	\flow{T'_p}{\ssp_p} = \LieEl'_p \ssp_p \, . 
\eeq
For the \rpo\ pair found by \rf{AvMeRoHo13}, 
$\LieEl'_p = \LieEl_\theta(\pi / 2) \LieEl_z (l'_p) $. 
For simplicity, we are going 
to treat this orbit as if its period were twice of its fundamental
period $T_p = 2 T'_p$, and its only symmetry is a streamwise 
shift since 
$\LieEl_p = (\LieEl'_p)^{2} = \LieEl_z (l_p = 2 l'_p)$. 

A \rpo\ with a one-parameter compact continuous symmetry defines a 
$2$-torus 
\beq
\{\LieEl_z (l) \flow{\zeit}{\ssp_p}\,|\,\zeit \in [0, T_p)\, , l \in [0, L) \}
\label{e-Torus}
\eeq 
in the \statesp , 
which is parametrized by shifts $l$ and time $\zeit$. 
Stability of a \rpo\ is determined 
by the eigenvalues of the Jacobian matrix
\beq
	J_p = \LieEl_p^{-1} \, d \flow{T_p}{\ssp_p} / d \ssp_p \, ,
	\label{e-Jac}
\eeq 
that are known as Floquet multipliers\rfp{DasBuch}. Since this Jacobian 
matrix
\refeq{e-Jac} is very large, in practice, we compute the leading Floquet 
multipliers $\Lambda_i$ and the corresponding Floquet vectors $V_i$ via 
Arnoldi iteration\rfp{Trefethen97}. The lower-branch \rpo\ of 
\rf{AvMeRoHo13} has only one positive real Floquet multiplier 
$\Lambda_1$ greater than one, 
two marginal 
$\Lambda_{2,3}\!=\!1$ multipliers corresponding to the disturbances as 
axial- and temporal-shifts; the rest $|\Lambda_{i>3}|$
of the multipliers have absolute
values smaller than one. Counting axial and 
temporal shift directions, the unstable manifold of this \rpo\ is 
three-dimensional. However, these marginal directions are of no 
dynamical importance and our next step is to cancel them.

\section{Reductions}
\label{s-sym}

For a $(1\!+\!1)$-dimensional partial differential equation
under a periodic boundary condition,\rf{BudCvi14} shows that 
the translation symmetry can be reduced by fixing the phase of 
the first Fourier mode to $0$
and this 
transformation can be interpreted as a ``slice'', that is a 
codimension-1 hyperplane 
where all symmetry-equivalent \statesp\ points
are represented by a single point. 
\rf{WiShCv15} adapted this
idea to pipe flow by taking a typical turbulent state as a template and
retaining only first Fourier mode components. 
Here, we take a much 
simpler approach and define a \textit{\fFslice} template $\slicep$ 
as the state vector corresponding to the three-dimensional field
\beq
	[\hat{u}', \hat{v}', \hat{w}'](z, \theta , r) 
        = J_0(\alpha r) \cos (2 \pi z / L) \,,
	\label{e-slicep}
\eeq
where $J_0$ is the zeroth Bessel function of the first kind and
$\alpha$ is chosen such that $J_0 (\alpha D/2) = 0$. For a given 
trajectory $\ssp (\zeit)$, we can now define a symmetry-reduced 
trajectory as 
\bea
	\sspRed (\zeit) &=& \LieEl_z (L \phi / 2 \pi  ) \ssp (\zeit) 
	\,, \, \mbox{where} \label{e-sspred} \\
	\phi (\zeit) &=& \arg (\inprod{\ssp(\zeit)}{\slicep} + i 
			       \inprod{\ssp(\zeit)}{\LieEl_z(- L/4) \slicep})
			       \,. \label{e-slicePhase} 
\eea
The \statesp\ 
vector pair $\slicep, \LieEl_z(- L/4) \slicep$ are orthogonal to 
each other and span a $2$-dimensional subspace.
The slice-fixing phase $\phi$ in \refeq{e-slicePhase} is 
the polar angle, when a state is projected onto this hyperplane. 
Transformation \refeq{e-sspred} fixes this angle to $0$, 
defining a unique symmetry reduced $\sspRed (\zeit)$ for all 
$\ssp (\zeit)$. 
We have chosen the $r$ dependence of the slice template in 
\refeq{e-slicep} to be a Bessel function because
of the cylindrical geometry. In practice, many other choices can be
equally good for the purpose of symmetry reduction. 

Our next step is to redefine the transformation \refeq{e-sspred}
as a slice, a codimension-1 half-hyperplane 
\beq
	\inprod{\sspRed(\zeit) - \slicep}{\sliceTan{}} = 0 \, , \quad
	\inprod{\groupTan (\sspRed)}{\sliceTan{}} > 0 \, , 
	\label{e-sliceCond}
\eeq
where 
$\groupTan (\slicep) = \Lg_z \slicep$ is the \textit{group tangent},
$\sliceTan{} = \groupTan (\slicep)$ is  the \textit{slice tangent}, 
$\Lg_z$ is the generator of infinitesimal 
translations, satisfying $\LieEl_z (l) = \exp \Lg_z l $. 
In our particular case,
$\Lg_z {\bf u} = - d {\bf u}/ dz$, hence 
$\sliceTan{} = \LieEl_z(- L/4) \slicep$. 
After defining the slice by \refeq{e-sliceCond}, 
one looks for the phases $\phi$
such that $\LieEl_z (L \phi / 2 \pi  ) \ssp (\zeit)$ satisfies 
\refeq{e-sliceCond}. 

At first sight, redefining the polar coordinate transformation
\refeq{e-sspred} as a slice \refeq{e-sliceCond} might seem 
as an over-complication, however, this reformulation provides 
some important tools. In particular, one can 
derive (see Appendix \ref{s-proj}) the projection operator 
\beq
	H (\sspRed) = 
	1 - \frac{\groupTan (\sspRed) \otimes 
   	    \sliceTan{}}{\inprod{\groupTan(\sspRed)}{\sliceTan{}}} 
   	\label{e-ProjSlice}
\eeq
that projects infinitesimal perturbations to $\sspRed$ from full
\statesp\ to the slice. With \refeq{e-sliceCond} and 
\refeq{e-ProjSlice}, we can now
reduce the torus \refeq{e-Torus} 
to a closed curve and define its stability. 
Let $\ssp_p$ be a point on a \rpo\ and
$V_i$ be the Floquet vectors computed at this point. 
If $\sspRed_p = g(L \phi_p / 2 \pi ) \ssp_p$ is the symmetry reduced 
\statesp\ point corresponding to $\ssp_p$, then the symmetry reduced
Floquet vectors are
$\hat{V}_i = H (\sspRed_p) g(L \phi_p / 2 \pi ) V_i$.
Note that $H (\sspRed) \groupTan (\sspRed) = 0$, thus 
the continuous symmetry direction is eliminated in the slice. 

It can be shown\rfp{DasBuch,guckb} that in the vicinity of a \po\ of a 
dynamical system, one can define a Poincar\'e map, which 
contains a fixed point, whose stability multipliers are equal
to the Floquet multipliers of the \po , except the marginal 
multiplier corresponding to the time direction that is 
eliminated by the Poincar\'e
section. Moreover, the stability 
eigenvectors in the Poincar\'e section can be obtained from Floquet
vectors by a projection. We define such a Poincar\'e section as
\beq
	\inprod{\sspRed_\PoincS - \sspRed_p}{\velRed (\sspRed_p)} = 0 
	\, , \quad 
	\inprod{\velRed (\sspRed_\PoincS)}{\velRed (\sspRed_p)} > 0 \, ,
	\label{e-Psect}
\eeq
where $\velRed (\sspRed) = H(\sspRed) \vel(\sspRed)$ and
$\vel(\ssp) = 
\lim_{\delta \zeit \rightarrow 0} 
\flow{\delta \zeit}{\ssp} / \delta \zeit$ is the 
\textit{\statesp\ velocity}. 
Con\-tin\-u\-ous-time symmetry-reduced flow 
$\sspRed (\zeit) = \flowRed{\zeit}{\sspRed(0)}$ induces a 
discrete-time dynamical system 
$\sspRed_\PoincS [n] = \flowP{n}{\sspRed_\PoincS[0]}$, where
$n$ counts the number of intersections of trajectories with the
Poincar\'e section as the discrete-time variable. Finally, we 
define the operator that reduces the infinitesimal perturbations
to $\sspRed_\PoincS$ from the symmetry-reduced \statesp\ to the 
Poincar\'e section as
\beq
	\ProjPsect(\sspRed_\PoincS) = 1 
			   - \frac{\velRed(\sspRed_\PoincS)  \otimes 
		               \velRed(\sspRed_p) }{
		               \inprod{\velRed(\sspRed_\PoincS)}{
				               \velRed(\sspRed_p)}} \, ,
	\label{e-ProjPoincare}
\eeq
which allows us to project the symmetry-reduced Floquet 
vectors $\hat{V}_i$ on to the Poincar\'e section as 
$\hat{V}_{i,\PoincS} = \ProjPsect(\sspRed_p) \hat{V}_i$.
Similar to \refeq{e-ProjSlice},
the projection operator \refeq{e-ProjPoincare} follows from
\refeq{e-Projection}.

%%%%%%%%%%%%%%%%%%%%%%%%%%%%%%%%%%%%%%%%%%%%%%%%%%%%%%%%%%%%%%%%%%%%%%%
\begin{figure}
	\centering
	  \includegraphics[width=0.5\textwidth]{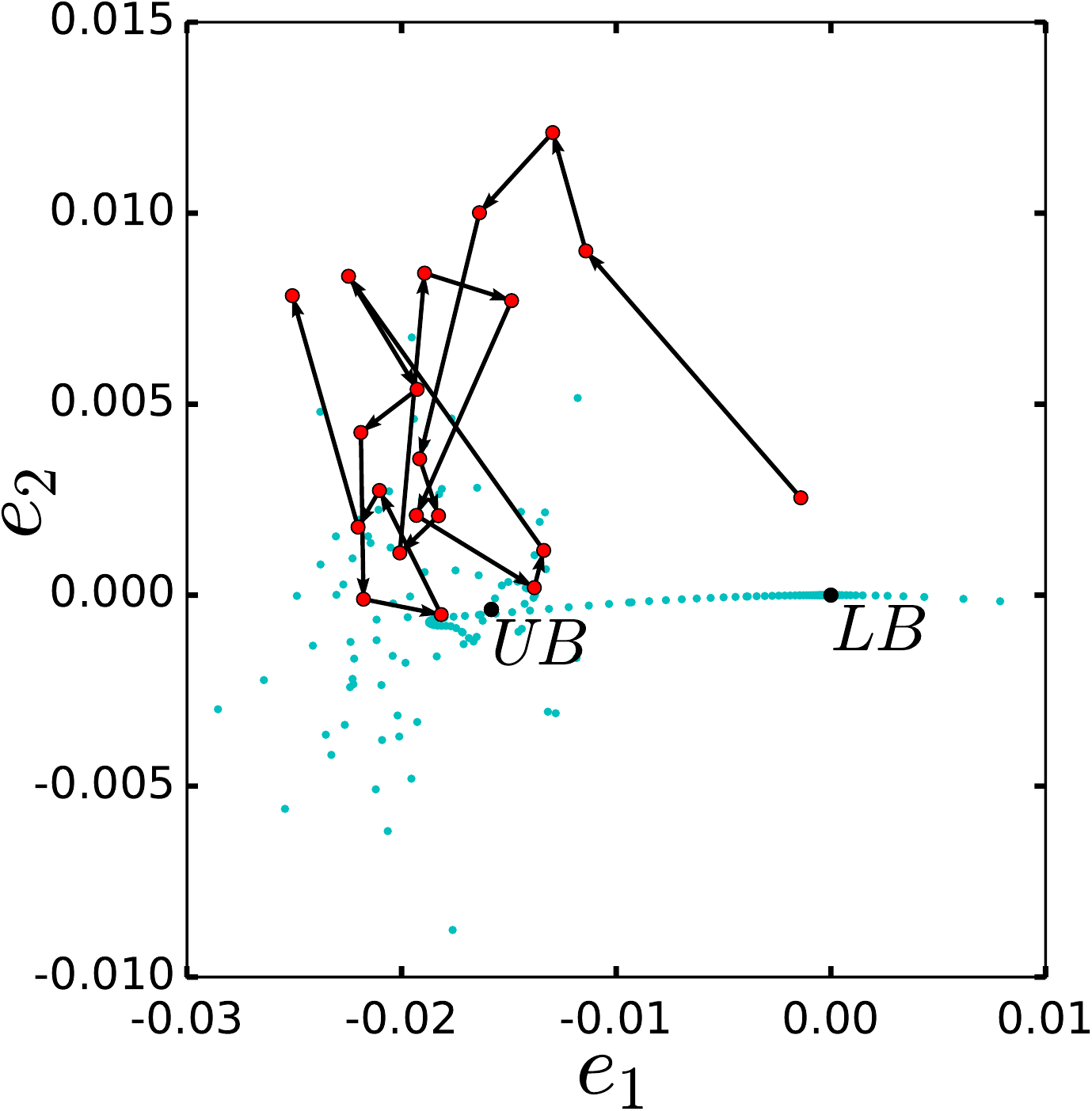} 
	\caption{\label{f-UnstMan}
		Unstable manifold of LB (cyan dots) and a small 
		perturbation (red circles)
		to the laminar flow that develops into a puff on the 
		Poincar\'e section \refeq{e-Psect}. 
		Direction of discrete time from $n=0$ 
		to $n=20$ is indicated by arrows for the perturbation.
		Intersection of \rpo s 
		LB and UB are marked black and annotated.
		Few points on the unstable manifold to the right 
		side of LB are shown as this direction consist of 
		laminarizing orbits.
	} 
\end{figure}
%%%%%%%%%%%%%%%%%%%%%%%%%%%%%%%%%%%%%%%%%%%%%%%%%%%%%%%%%%%%%%%%%%%%%%

\section{The unstable manifold}
\label{s-man}

\rf{BudCvi15}, demonstrate the computation of one- and
two-dimensional unstable manifolds of \rpo s 
on Poincar\'e sections 
for the \KS\ system. The main idea is to select trajectories 
that approximately cover the linear unstable manifold, hence their
forward integration approximately covers the nonlinear unstable 
manifold; see \rf{BudCvi15} for details. Here, we are interested in 
computing the unstable manifold of the lower-branch \rpo\ (LB) of 
\rf{AvMeRoHo13} at $\Reynolds = 1700$. At this \Reynolds, the only
unstable Floquet multiplier of LB is $\Lambda_1 = 2.8291523$.
Initial conditions on the Poincar\'e section that approximately cover
the associated unstable manifold are
\beq
	\sspRed_\PoincS (\delta) = \sspRed_{p, \PoincS} \pm 
							   \epsilon \Lambda^{\delta}_1
							   \hat{V}_{1, \PoincS}\, , \quad
\eeq
where $\delta \in [0,1)$ and $\epsilon$ is a small constant. 
We set $\epsilon = 10^{-4}$ and discretize $\delta$ by taking 
$9$ equidistant points in $[0, 1)$. 

\refFig{f-UnstMan} shows
the unstable manifold approximated this way. Here, we show the first
$20$ intersections of each trajectory with the Poincar\'e section,
projected onto bases formed by the unstable Floquet vector $\hat{V}_1$ 
and the least stable Floquet vector $\hat{V}_4$ of LB, orthonormalized
by the Gram-Schmidt procedure, namely
\(
e_1 = \langle \sspRed_\PoincS - \sspRed_p, \hat{V}_{1,\bot}\rangle
,\,  
e_2 = \langle\sspRed_\PoincS - \sspRed_p,
	  \hat{V}_{4,\bot}\rangle
\), where subscript $\bot$ implies that vectors are orthonormalized.
We observed that as the trajectories 
leave the neighbourhood of LB (the origin of \reffig{f-UnstMan}) 
they approach the UB. This indicates the existence of
a heteroclinic connection between the two. Another trajectory
shown on \reffig{f-UnstMan} is an initially small perturbation 
developing into a puff. 
We generated this perturbation by scaling down a typical puff state
and found that it approached the UB before redeveloping into a puff.

\begin{figure}
	\centering
	(a) \includegraphics[width=0.3\textwidth]{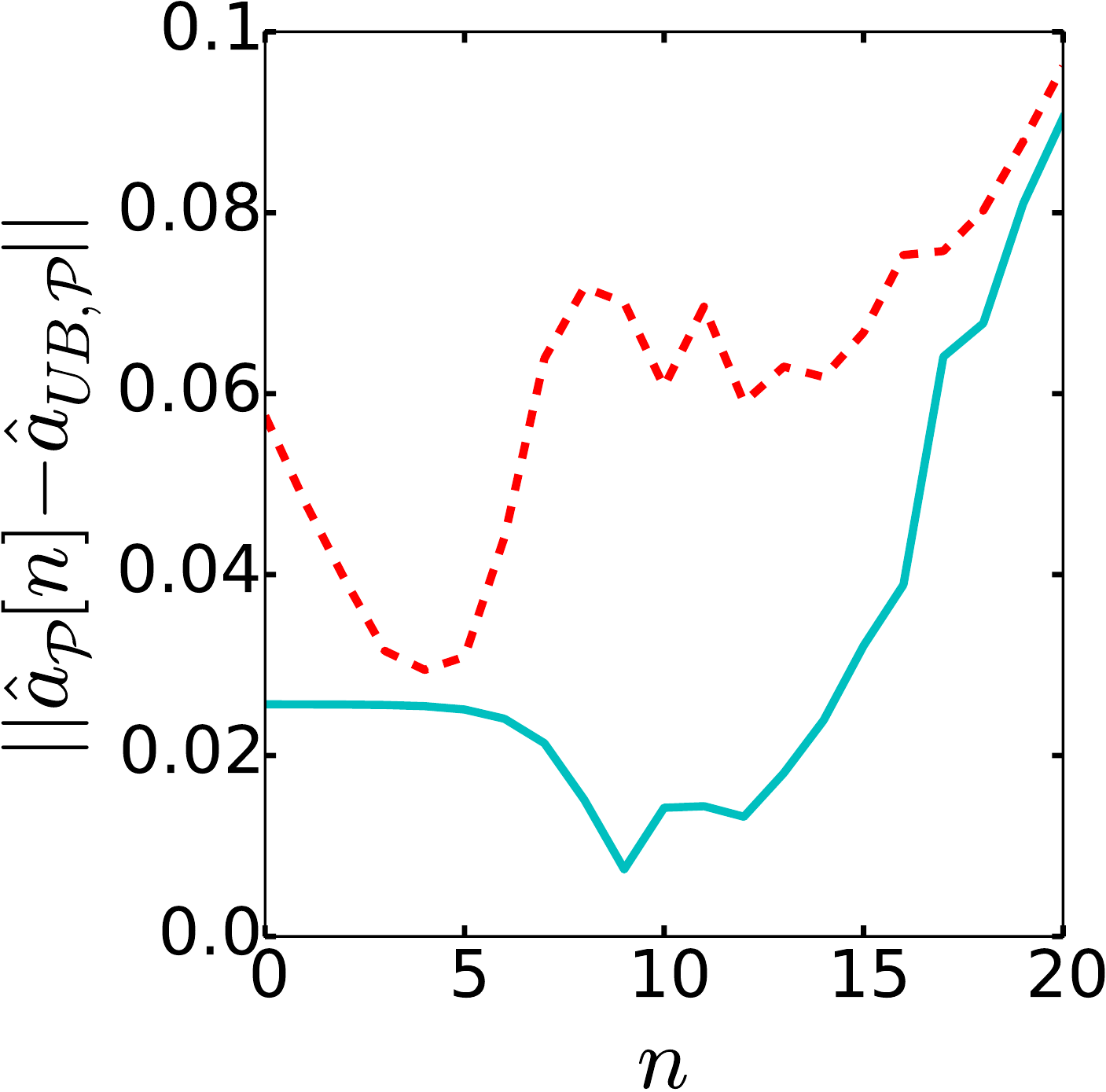} \qquad
	(b) \includegraphics[width=0.3\textwidth]{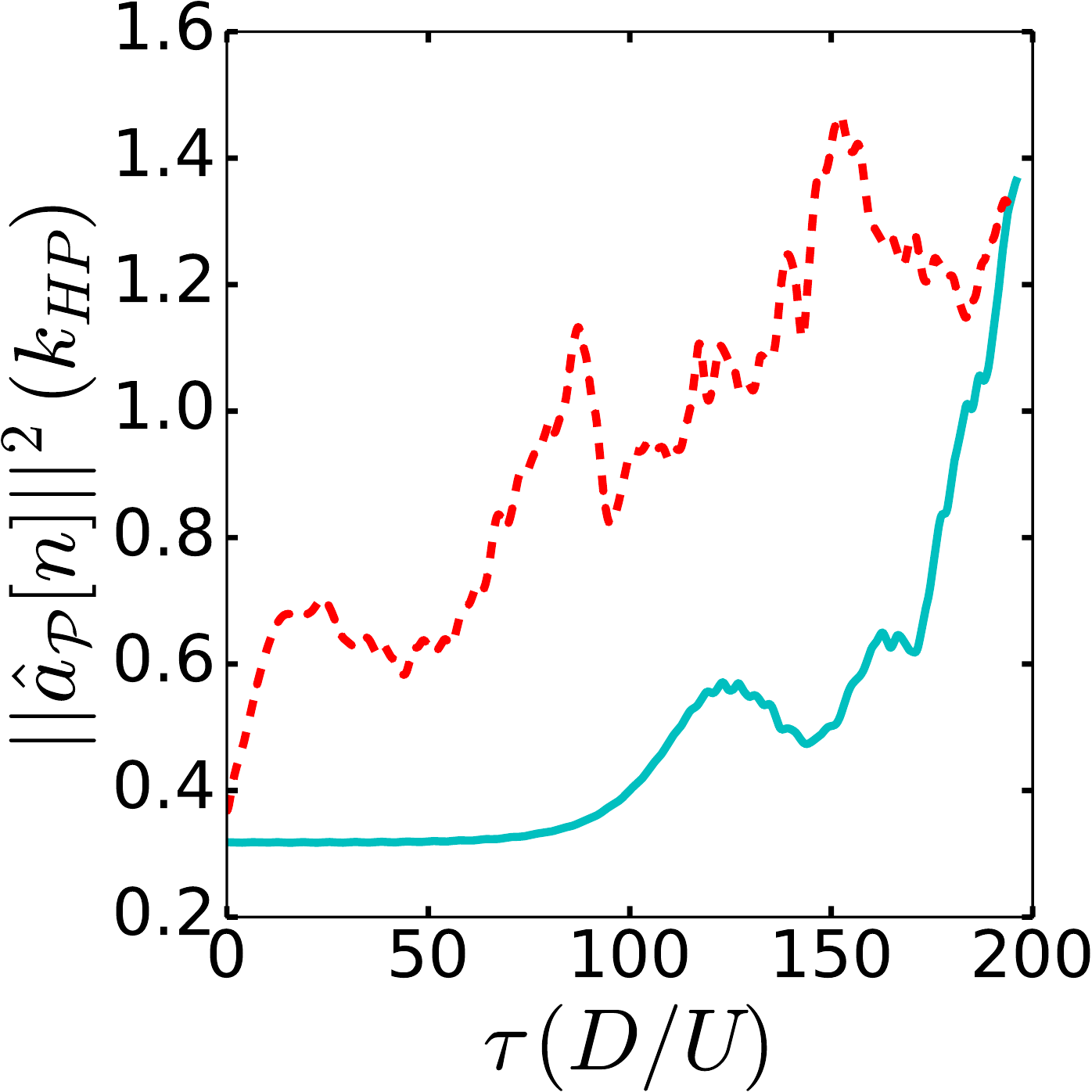} 
	\caption{\label{f-tSeries}
		(a) $L_2$ Distance from UB on the Poincar\'e 
		section for $20$ returns, 
		(b) Time-evolution of kinetic energy in units
		of the kinetic energy $k_{HP}$ of the 
		laminar flow. Cyan (solid): A trajectory on the 
		unstable manifold of LB, red (dashed): puff transition
		initiated by a small perturbation to the laminar state.
	} 
\end{figure}

In order to confirm apparent approaches to the UB, we measured the 
trajectories' distance from it on the Poincar\'e section. 
\refFig{f-tSeries}(a) shows the distance of an orbit on the 
unstable manifold of LB and the perturbation we show on 
\reffig{f-UnstMan} from the UB. For both trajectories, we see a 
clear initial drop before they move away following the 
unstable manifold of the UB. 
Note that some of the intersections of initially small
	  perturbation's trajectory with the Poincar\'e section appear
	  closer to the UB than its closest approach ($n = 4$)
	  in \reffig{f-UnstMan}. This is an artefact of low-dimensional
	  projection from an infinite dimensional Poincar\'e section.
For further comparison, we show the time-evolution
of turbulent kinetic energy for both trajectories on 
\reffig{f-tSeries}(b).
Note for the trajectory on the unstable manifold that before the 
kinetic energy goes up to puff levels ($\sim 1.4 k_{HP}$) it 
oscillates around $0.5 k_{HP}$ during $\zeit \in (125,175) (D/U)$.
Similarly, for the small perturbation, after initial increase, 
the kinetic energy stays close to $0.6 k_{HP}$ during 
$\zeit \in (25,75) (D/U)$ before further increasing to puff levels.
Both episodes corresponds to trajectories' approach to UB.
Note that the 
time interval shown in \reffig{f-tSeries}(a) is not necessarily the
same for each orbit, nor are the discrete time intervals equal 
to each other. 
For the perturbation, the interval $n \in [0, 20]$ 
to $\tau \in (0,146.25)\,(D/U)$, 
whereas for the trajectory on the unstable manifold, 
it corresponds to $\tau \in (0, 193.67)\,(D/U)$.
%%%%%%%%%%%%%%%%%%%%%%%%%%%%%%%%%%%%%%%%%%%%%%%%%%%%%%%%%%%%%%%%%%%%%%%
%%%%%%%%%%%%%%%%%%%%%%%%%%%%%%%%%%%%%%%%%%%%%%%%%%%%%%%%%%%%%%%%%%%%%%
%%%%%%%%%%%%%%%%%%%%%%%%%%%%%%%%%%%%%%%%%%%%%%%%%%%%%%%%%%%%%%%%%%%%%%%
\begin{figure}
	\centering
	(a) \includegraphics[width=0.95\textwidth]{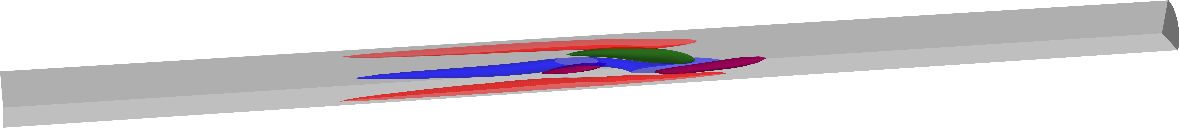}  \\
	(b) \includegraphics[width=0.95\textwidth]{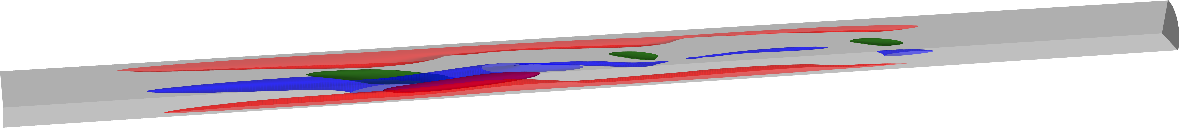}  \\
	(c) \includegraphics[width=0.95\textwidth]{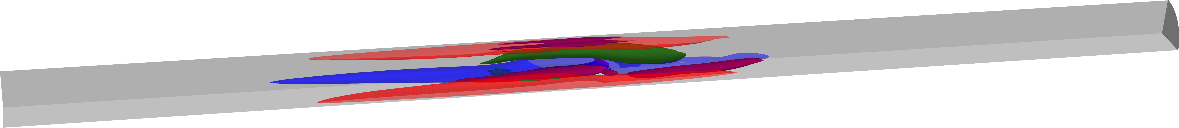}  \\
	(d) \includegraphics[width=0.95\textwidth]{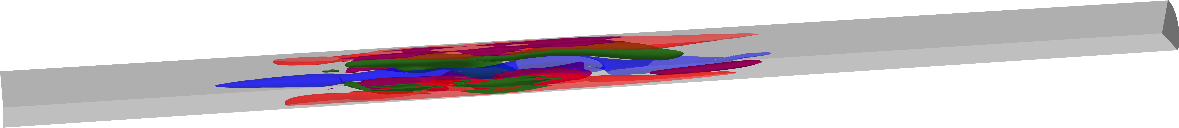}  \\
	(e) \includegraphics[width=0.95\textwidth]{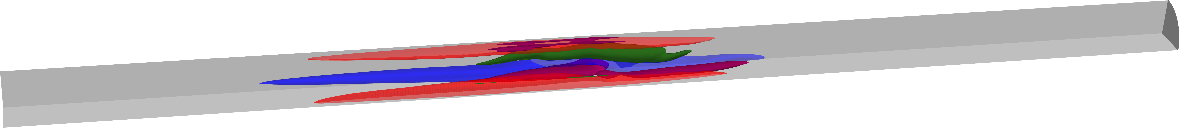} 	   \\
	(f) \includegraphics[width=0.95\textwidth]{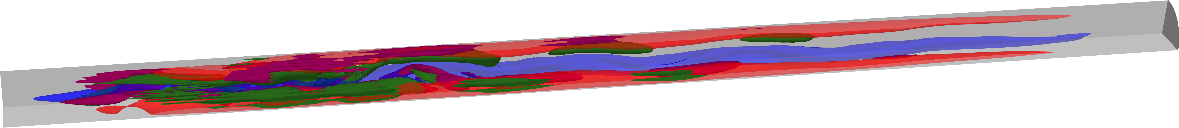} 
	\caption{\label{f-flowStruct}
		isosurfaces of streamwise vorticity at 
        $\omega_z = \pm 0.5 U / D$ (pink and green) 
        and streamwise velocity at
        $u = \pm 0.1 U$ (red and blue)
		for the trajectory on LB's unstable manifold at 
		$n = 0, 9, 20$ (a, c, f), 
		for the small perturbation at 
		$n = 0, 4, 20$ (b, d) and for UB (e).
        Shown here is the pipe region
        $z \in [5, 20] D\, , \theta \in [0, \pi/2]$ 
        since rest of the structures
        can be obtained from reflection and 
        rotation by $\pi$ symmetries
        that are present.
	}
\end{figure}
%%%%%%%%%%%%%%%%%%%%%%%%%%%%%%%%%%%%%%%%%%%%%%%%%%%%%%%%%%%%%%%%%%%%%%
For further comparison, we visualized streamwise velocity and vorticity 
isosurfaces 
of these orbits and the UB on 
\reffig{f-flowStruct}. 
At the closest approach ($n=9$, \reffig{f-flowStruct} (c)) 
the resemblance of flow structures of the 
unstable manifold with those of UB 
(\reffig{f-flowStruct} (e)) is very close. 
While not as 
dramatic, at closest approach 
($n=4$, \reffig{f-flowStruct} (d))
of the initially (\reffig{f-flowStruct} (b)) 
small perturbation's to UB, we also see structural resemblance. 
The same isosurfaces for the LB are visually indistinguishable 
from the initial point ($n=0$, \reffig{f-flowStruct} (a)) on the 
unstable manifold, hence not separately shown in 
\reffig{f-flowStruct}. Note that puffs are structurally 
(\reffig{f-flowStruct} (f)) much more complicated
than $LB$ and $UB$.

In order to compare sizes of turbulent structures, we plotted the
kinetic
energy of fluctuations as a function of axial position,
\ie\ 
$k(z) = \frac{1}{2} 
		\int_0^{D/2} \int_0^{2 \pi} 
		{\bf u} \cdot {\bf u} \, r dr d\theta $,
for the trajectory 
on the unstable manifold of LB. \refFig{f-sizes} (a) shows the
time interval $n \in [0, 9]$, during which the orbit leaves the 
neighbourhood of LB and approaches to UB. 
During the time-interval
$n \in [10, 20]$ shown in \reffig{f-sizes} (b), the orbit moves away 
from UB and becomes a puff. 

\section{Conclusion and outlook}
\label{s-con}
%%%%%%%%%%%%%%%%%%%%%%%%%%%%%%%%%%%%%%%%%%%%%%%%%%%%%%%%%%%%%%%%%%%%%%%
\begin{figure}
	\centering
	(a) \includegraphics[width=0.45\textwidth]{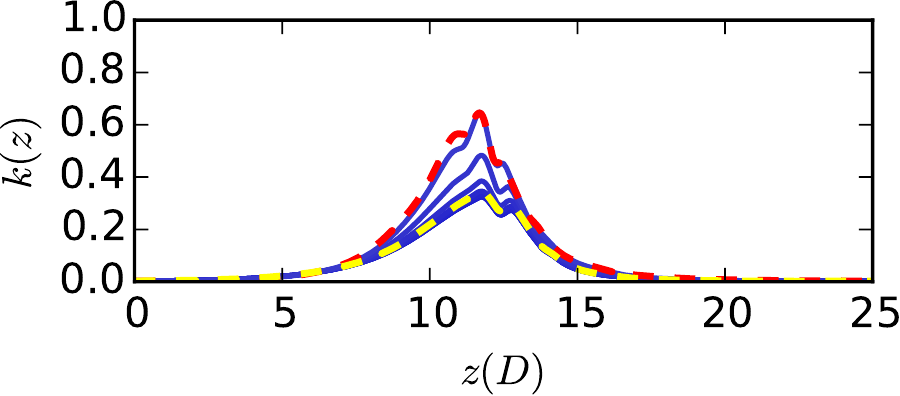}  
	(b) \includegraphics[width=0.45\textwidth]{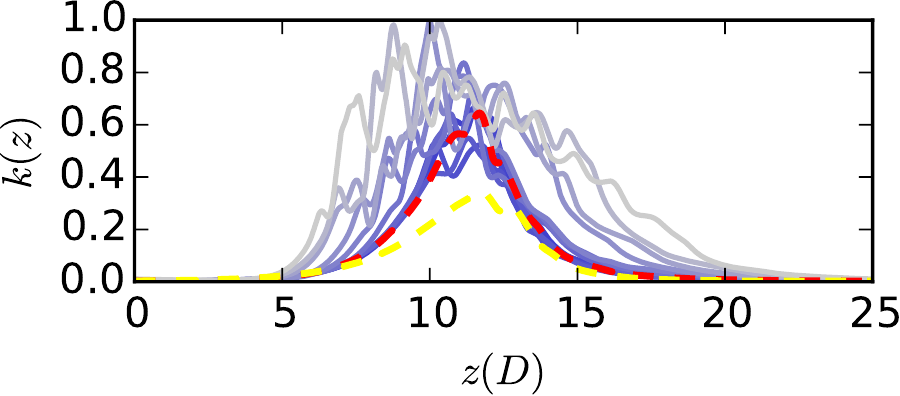}  
	\caption{\label{f-sizes}
		Fluctuating kinetic energy $k(z)$ (arbitrary units) contained at an 
        axial location $(z)$ for $n = 0, 1, \ldots , 9$ and 
        $n = 10, 11, \ldots, 20$ (b). Darkest shade of
        blue corresponds to initial time $n=0$, 
        and the plotting colour becomes lighter 
        as time increases. Dashed yellow and red  
        curves correspond to the LB and the UB.
        $k(z)$ values are normalized by their maximum.
	}
\end{figure}
%%%%%%%%%%%%%%%%%%%%%%%%%%%%%%%%%%%%%%%%%%%%%%%%%%%%%%%%%%%%%%%%%%%%%%
In this paper, we presented strong numerical evidence of a heteroclinic 
connection between two spatially localized \rpo s of pipe flow. 
This is a remarkable example of structural stability in
a dynamical system since these orbits born of a saddle-node 
bifurcation at $\Reynolds \approx 1430$ yet their dynamical 
connection persists at $\Reynolds = 1700$.
We repeated our unstable manifold calculation at 
		  $\Reynolds = 1900$ and obtained a qualitatively similar 
		  picture.

Previous studies \rfp{GHCV08,VaKa11} reported homoclinic and 
heteroclinic connections in Couette flow, which can accommodate 
	equilibria and periodic orbits without spatial drifts.
This is not the case for pressure-driven flows since
all invariant solutions except the laminar equilibrium have non-zero 
streamwise drifts. In this regard, the current work fills an important 
technical gap and provides a new set of tools for study of 
pressure-driven flows.
While, the methods we used here appeared in 
different publications\rfp{BudCvi14,DCTSCD14,BudCvi15},
where they are applied to much lower-dimensional systems, 
this is their first application to the full \NSe. 

While \rf{AvMeRoHo13}
shows that the chaotic motion emerges from the bifurcations of 
the UB,
as the Reynolds number is increased from $1430$ to $1545$,
the role of UB for transition is not obvious at $\Reynolds = 1700$.
Our result shows that far from the bifurcation point, the UB takes 
the role of mediating the transition. To the best of our knowledge,
this is the first example of an upper-branch solution in a shear flow 
that lies between laminar and the turbulent parts of the \statesp , 
similar to lower-branch. Note that UB, whose kinetic 
energy swings around $k = 0.5 k_{HP}$, is energetically separated 
from turbulent puffs which have typical kinetic energies 
$k \approx 1.2 k_{HP}$. 
This is also clearly visible from the axial distribution of kinetic 
energies on \reffig{f-sizes} and flow structures on
\reffig{f-flowStruct}, where the UB clearly has a much simpler 
structure than puffs.

The time evolution of the spatial distribution of the 
kinetic energy on \reffig{f-sizes} suggests a two-stage transition 
scenario, where the spatial complexity of a puff forms as
trajectories follow
the unstable manifold of UB (\reffig{f-sizes} (b)). 
\rf{RiMeAv16} recently reported a detailed numerical study, where they
used turbulent kinetic energy and pressure gradient as
indicators to support the hypothesis that spatial complexity in this 
system arises as different chaotic regions in the \statesp\ 
merge with the neighbourhood of the UB. These observations along 
with ours motivate a detailed study of UB's unstable manifold
in order to understand spatial expansion of chaotic spots. 
The tools we introduce here can be useful for such a study.

\begin{acknowledgments}
We are indebted to
Ashley P. Willis for making his DNS code and invariant solutions
available on \texttt{openpipeflow.org},
to Predrag Cvitanovi\'c,
   Marc Avila, and
   Genta Kawahara
for fruitful discussions, 
and to Yohann Duguet for his critical reading of an early 
      version of the manuscript. 
This research was supported in part by the National Science
Foundation under Grant No. NSF PHY11-25915.
\end{acknowledgments}

\begin{appendices}	
\section{Projection operator}
\label{s-proj}
	\begin{figure}
		\begin{center}
			\setlength{\unitlength}{0.3\textwidth}
			\begin{picture}(1,1.02633025)%
			\put(0,0){\includegraphics[width=\unitlength,page=1]{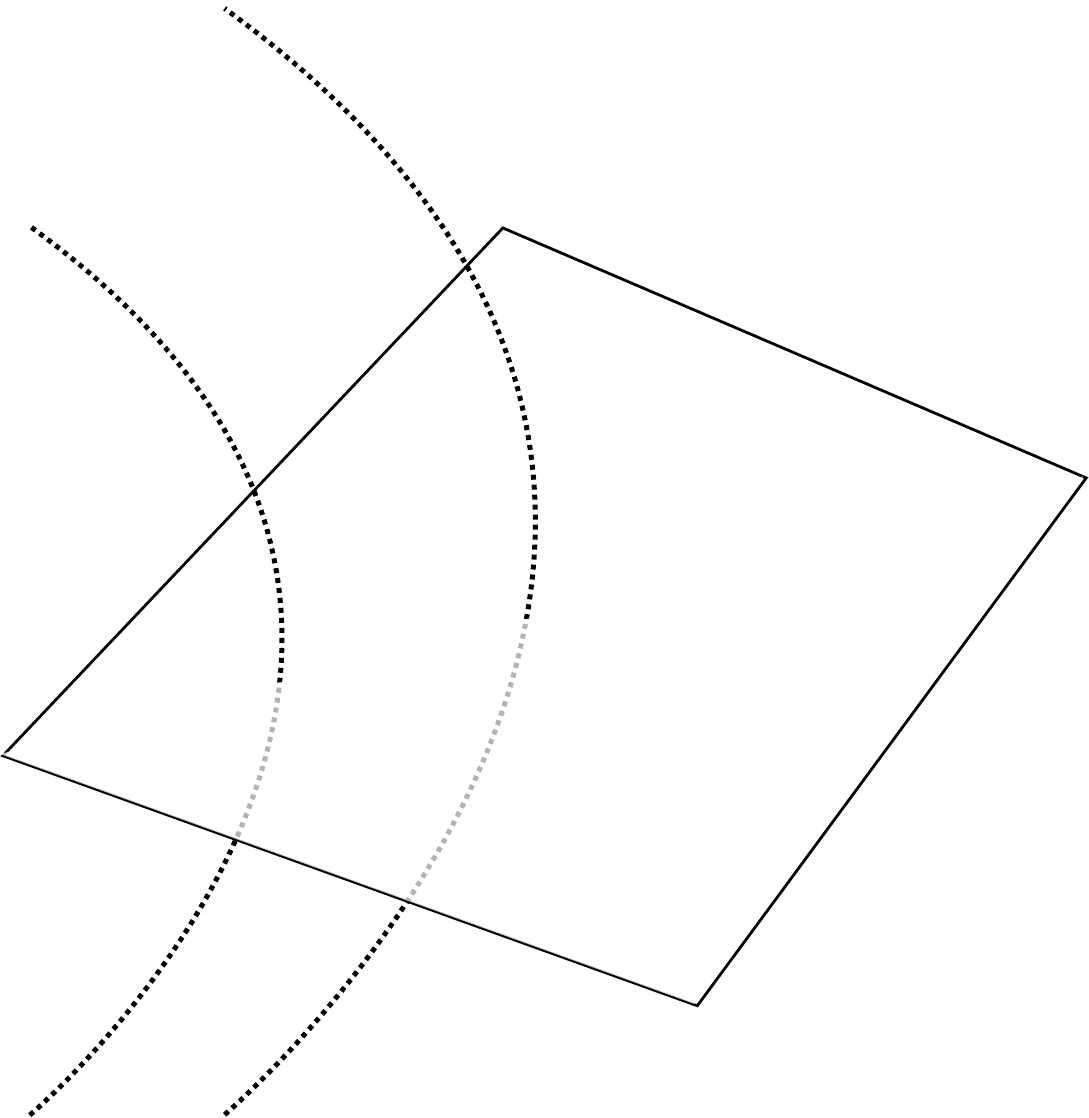}}%
			\put(0.63,0.6){\color[rgb]{0,0,0}\makebox(0,0)[lt]{\begin{minipage}{0.35522149\unitlength}\raggedright $U(\sspRed) = 0$ \end{minipage}}}%
			\put(0,0){\includegraphics[width=\unitlength,page=2]{transTang.pdf}}%
			\put(0.08,0.74371883){\color[rgb]{0,0,0}\makebox(0,0)[lt]{\begin{minipage}{0.07656645\unitlength}\raggedright $\ssp$ \end{minipage}}}%
			\put(0,0){\includegraphics[width=\unitlength,page=3]{transTang.pdf}}%
			\put(0.26798256,1.03540051){\color[rgb]{0,0,0}\makebox(0,0)[lt]{\begin{minipage}{0.29350471\unitlength}\raggedright $\ssp + \delta \ssp$ \end{minipage}}}%
			\put(0,0){\includegraphics[width=\unitlength,page=4]{transTang.pdf}}%
			\put(0.2,0.4){\color[rgb]{0,0,0}\makebox(0,0)[lt]{\begin{minipage}{0.07656645\unitlength}\raggedright $\sspRed$\end{minipage}}}%
			\put(0.49403588,0.44){\color[rgb]{0,0,0}\makebox(0,0)[lt]{\begin{minipage}{0.29350471\unitlength}\raggedright $\sspRed + \delta \sspRed$ \end{minipage}}}%
			\put(0.0,0.62){\color[rgb]{0,0,0}\makebox(0,0)[lt]{\begin{minipage}{0.07656645\unitlength}\raggedright $\flow{\gSpaceRed}{\ssp}$ \end{minipage}}}%
			\put(0.43205352,0.9004977){\color[rgb]{0,0,0}\makebox(0,0)[lt]{\begin{minipage}{0.40835438\unitlength}\raggedright $\flow{\gSpaceRed + \delta \gSpaceRed}{\ssp + \delta \ssp}$ \end{minipage}}}%
			\end{picture}%
		\end{center}
		\caption{\label{f-hyperplane} Schematic illustration of the transformation
			of $\ssp$ and a small perturbation $\delta \ssp$ to it by the nonlinear 
			group action $\flow{\phi}{\ssp}$ onto the hypersurface $U (\sspRed) = 0$.}
	\end{figure}
	The projection operations \refeq{e-ProjSlice} and 
	\refeq{e-ProjPoincare} follow from the 
	same geometrical principle, which 
	we are going to derive here. Let $\flow{\phi}{\ssp}$ be the 
	nonlinear semi-group action that transforms the 
	\stateDsp\ vector $\ssp$ 
	according to the parameter $\phi$ as 
	\beq
	\ssp' = \flow{\gSpace}{\ssp}\,,\quad \gSpace \in [0, \gSpace_{max})
	\eeq
	and let $U(\ssp)$
	be a scalar-valued function of $\ssp$ that
	defines a codimension-1 hypersurface $U(\sspRed) = 0$ in the \stateDsp\ such that at 
	$\sspRed = \flow{\gSpaceRed}{\ssp}$ a semi-group orbit of $\ssp$ 
	intersects this hypersurface transversally as illustrated in 
	\reffig{f-hyperplane}. Now let us consider a small perturbation 
	$\delta \ssp$ to $\ssp$ and its transformation onto this hypersurface: 
	\beq
	\sspRed + \delta \sspRed 
	= \flow{\gSpaceRed + \delta \gSpaceRed}{\ssp + \delta \ssp} \, .
	\label{e-PerturbRed}
	\eeq
	Taylor expanding the RHS to linear order in $\delta \gSpaceRed$ 
	and $\delta \ssp$ we obtain
	\beq
	\delta \sspRed = 
	\partial_\gSpace 
	\flow{\gSpace}{\ssp}|_{\gSpace = \gSpaceRed} \delta \gSpaceRed
	+ \left.\frac{d \flow{\gSpace}{\ssp'}}{d \ssp'}\right|_{\ssp' = \ssp}
	\delta \ssp \, . \label{e-Taylor1}
	\eeq
	Our goal is to find an expression for $\delta \sspRed$, however,
	\refeq{e-Taylor1} gives us one condition with two unknowns 
	$\delta \sspRed$ and $\delta \gSpaceRed$.
	The second condition comes from the fact that 
	\refeq{e-PerturbRed} also
	satisfies the hypersurface equation, \ie\ 
	$U (\sspRed + \delta \sspRed) = 0 $. Taylor expansion to the linear
	order yields
	\bea
	\inprod{
		\left.\frac{d U(\ssp)}{d \ssp}\right|_{\ssp = \sspRed} }
	{\delta \sspRed} &=& 0 \, , \continue
	\inprod
	{\left.\frac{d U(\ssp)}{d \ssp}\right|_{\ssp = \sspRed} }
	{\partial_\gSpace 
		\flow{\gSpace}{\ssp}|_{\gSpace = \gSpaceRed}} \delta \gSpaceRed
	+ 
	\inprod
	{\left.\frac{d U(\ssp)}{d \ssp}\right|_{\ssp = \sspRed} }
	{\left.\frac{d \flow{\gSpace}{\ssp'}}{d \ssp'}\right|_{\ssp' = \ssp}
		\delta \ssp}
	&=& 0 \, , \label{e-Taylor2}
	\eea
	where in the second step we inserted \refeq{e-Taylor1} for 
	$\delta \sspRed$. Solving \refeq{e-Taylor2} for $\delta \gSpaceRed$
	and inserting its expression into \refeq{e-Taylor1} we find
	\beq
	\delta \sspRed = 
	\left.\frac{d \flow{\gSpace}{\ssp'}}{d \ssp'}\right|_{\ssp' = \ssp}
	\delta \ssp 
	- \partial_\gSpace 
	\flow{\gSpace}{\ssp}|_{\gSpace = \gSpaceRed} 
	\frac{\inprod
		{\left.\frac{d U(\ssp)}{d \ssp}\right|_{\ssp = \sspRed} }
		{\left.\frac{d \flow{\gSpace}{\ssp'}}{d \ssp'}\right|_{\ssp' = \ssp}
			\delta \ssp}}{
		\inprod
		{\left.\frac{d U(\ssp)}{d \ssp}\right|_{\ssp = \sspRed} }
		{\partial_\gSpace 
			\flow{\gSpace}{\ssp}|_{\gSpace = \gSpaceRed}}} \, , 
	\eeq
	which we can rewrite as
	\beq
	\delta \sspRed = 
	\left( \textbf{1} -  
	\frac{\partial_\gSpace 
		\flow{\gSpace}{\ssp}|_{\gSpace = \gSpaceRed} 
		\otimes 
		\left.\frac{d U(\ssp)}{d \ssp}\right|_{\ssp = \sspRed}}{
		\inprod
		{\left.\frac{d U(\ssp)}{d \ssp}\right|_{\ssp = \sspRed} }
		{\partial_\gSpace 
			\flow{\gSpace}{\ssp}|_{\gSpace = \gSpaceRed}}}
	\right)
	\left.\frac{d \flow{\gSpace}{\ssp'}}{d \ssp'}\right|_{\ssp' = \ssp}
	\delta \ssp \, , \label{e-Projection}
	\eeq
	where $\otimes$ denotes the outer product. 
	Both projection operators \refeq{e-ProjSlice} and \refeq{e-ProjPoincare} 
	can be obtained
	from \refeq{e-Projection} by substitutions 
	$(l, \LieEl(l) \ssp) \rightarrow (\gSpace, \flow{\gSpace}{\ssp})$ and
	$(\zeit, \flow{\zeit}{\ssp}) \rightarrow (\gSpace, \flow{\gSpace}{\ssp})$
	respectively.
\end{appendices}

 \bibliographystyle{jfm}
 \bibliography{../bibtex/neubauten.bib}

\begin{thebibliography}{26}
\expandafter\ifx\csname natexlab\endcsname\relax\def\natexlab#1{#1}\fi

\bibitem[Avila {\em et~al.\/}(2011)Avila, Moxey, de~Lozar, Avila, Barkley \&
  Hof]{AMdABH11}
{\sc Avila, K., Moxey, D., de~Lozar, A., Avila, M., Barkley, D. \& Hof, B.}
  2011 The onset of turbulence in pipe flow. {\em Science\/} {\bf 333},
  192--196.

\bibitem[Avila {\em et~al.\/}(2013)Avila, Mellibovsky, Roland \&
  Hof]{AvMeRoHo13}
{\sc Avila, M., Mellibovsky, F., Roland, N. \& Hof, B.} 2013
  Streamwise-localized solutions at the onset of turbulence in pipe flow. {\em
  Phys. Rev. Lett.\/} {\bf 110}, 224502.

\bibitem[Budanur \& Cvitanovi\'c(2015)]{BudCvi15}
{\sc Budanur, N.~B. \& Cvitanovi\'c, P.} 2015 Unstable manifolds of relative
  periodic orbits in the symmetry-reduced state space of the
  {Kuramoto-Sivashinsky} system. {\em J. Stat. Phys.\/} .

\bibitem[Budanur {\em et~al.\/}(2015)Budanur, Cvitanovi\'c, Davidchack \&
  Siminos]{BudCvi14}
{\sc Budanur, N.~B., Cvitanovi\'c, P., Davidchack, R.~L. \& Siminos, E.} 2015
  Reduction of the {SO(2)} symmetry for spatially extended dynamical systems.
  {\em Phys. Rev. Lett.\/} {\bf 114}, 084102.

\bibitem[Cvitanovi{\'c} {\em et~al.\/}(2015)Cvitanovi{\'c}, Artuso, Mainieri,
  Tanner \& Vattay]{DasBuch}
{\sc Cvitanovi{\'c}, P., Artuso, R., Mainieri, R., Tanner, G. \& Vattay, G.}
  2015 {\em Chaos: Classical and Quantum\/}. Copenhagen: Niels Bohr Inst., {\tt
  ChaosBook.org}.

\bibitem[Ding {\em et~al.\/}(2016)Ding, Chat\'e, Cvitanovi\'c, Siminos \&
  Takeuchi]{DCTSCD14}
{\sc Ding, X., Chat\'e, H., Cvitanovi\'c, P., Siminos, E. \& Takeuchi, K.~A.}
  2016 Estimating the dimension of the inertial manifold from unstable periodic
  orbits. {\em Phys. Rev. Lett.\/} {\bf 117}, 024101.

\bibitem[Duguet {\em et~al.\/}(2008)Duguet, Willis \& Kerswell]{duguet07}
{\sc Duguet, Y., Willis, A.~P. \& Kerswell, R.~R.} 2008 Transition in pipe
  flow: the saddle structure on the boundary of turbulence. {\em J. Fluid
  Mech.\/} {\bf 613}, 255--274, \arXiv{0711.2175}.

\bibitem[Faisst \& Eckhardt(2003)]{FE03}
{\sc Faisst, H. \& Eckhardt, B.} 2003 Traveling waves in pipe flow. {\em Phys.
  Rev. Lett.\/} {\bf 91}, 224502.

\bibitem[Guckenheimer \& Holmes(1983)]{guckb}
{\sc Guckenheimer, J. \& Holmes, P.} 1983 {\em Nonlinear Oscillations,
  Dynamical Systems, and Bifurcations of Vector Fields\/}. New York: Springer.

\bibitem[Halcrow {\em et~al.\/}(2009)Halcrow, Gibson, Cvitanovi{\'c} \&
  Viswanath]{GHCV08}
{\sc Halcrow, J., Gibson, J.~F., Cvitanovi{\'c}, P. \& Viswanath, D.} 2009
  Heteroclinic connections in plane {Couette} flow. {\em J. Fluid Mech.\/} {\bf
  621}, 365--376.

\bibitem[Hopf(1948)]{hopf48}
{\sc Hopf, E.} 1948 A mathematical example displaying features of turbulence.
  {\em Commun. Pure Appl. Math.\/} {\bf 1}, 303--322.

\bibitem[Jim{\'e}nez \& Moin(1991)]{JM91}
{\sc Jim{\'e}nez, J. \& Moin, P.} 1991 The minimal flow unit in near-wall
  turbulence. {\em J. Fluid Mech.\/} {\bf 225}, 213--240.

\bibitem[Lemoult {\em et~al.\/}(2016)Lemoult, Shi, Avila, Jalikop, Avila \&
  Hof]{LSAJAH16}
{\sc Lemoult, G., Shi, L., Avila, K., Jalikop, S.~V., Avila, M. \& Hof, B.}
  2016 Directed percolation phase transition to sustained turbulence in
  {Couette} flow. {\em Nat Phys\/} {\bf 12}, 254--258.

\bibitem[Nagata(1990)]{N90}
{\sc Nagata, M.} 1990 Three-dimensional finite-amplitude solutions in plane
  {Couette} flow: {Bifurcation} from infinity. {\em J. Fluid Mech.\/} {\bf
  217}, 519--527.

\bibitem[Ritter {\em et~al.\/}(2016)Ritter, Mellibovsky \& Avila]{RiMeAv16}
{\sc Ritter, P., Mellibovsky, F. \& Avila, M.} 2016 Emergence of
  spatio-temporal dynamics from exact coherent solutions in pipe flow. {\em New
  J. Phys.\/} {\bf 18}, 083031.

\bibitem[Schneider {\em et~al.\/}(2007)Schneider, Eckhardt \&
  Yorke]{SchEckYor07}
{\sc Schneider, T.~M., Eckhardt, B. \& Yorke, J.} 2007 Turbulence, transition,
  and the edge of chaos in pipe flow. {\em Phys. Rev. Lett.\/} {\bf 99},
  034502.

\bibitem[Schneider {\em et~al.\/}(2008)Schneider, Gibson, Lagha, Lillo \&
  Eckhardt]{SGLDE08}
{\sc Schneider, T.~M., Gibson, J.~F., Lagha, M., Lillo, F.~D. \& Eckhardt, B.}
  2008 Laminar-turbulent boundary in plane {Couette} flow. {\em Phys. Rev.
  E.\/} {\bf 78}, 037301, \arXiv{0805.1015}.

\bibitem[Skufca {\em et~al.\/}(2006)Skufca, Yorke \& Eckhardt]{SYE05}
{\sc Skufca, J.~D., Yorke, J.~A. \& Eckhardt, B.} 2006 {Edge of Chaos} in a
  parallel shear flow. {\em Phys. Rev. Lett.\/} {\bf 96}, 174101.

\bibitem[Toh \& Itano(2003)]{TI03}
{\sc Toh, S. \& Itano, T.} 2003 A periodic-like solution in channel flow. {\em
  J. Fluid Mech.\/} {\bf 481}, 67--76.

\bibitem[Trefethen \& Bau(1997)]{Trefethen97}
{\sc Trefethen, L.~N. \& Bau, D.} 1997 {\em Numerical Linear Algebra\/}. SIAM.

\bibitem[van Veen \& Kawahara(2011)]{VaKa11}
{\sc van Veen, L. \& Kawahara, G.} 2011 Homoclinic tangle on the edge of shear
  turbulence. {\em Phys. Rev. Lett.\/} {\bf 107}, 114501.

\bibitem[Waleffe(1998)]{W98}
{\sc Waleffe, F.} 1998 Three-dimensional coherent states in plane shear flows.
  {\em Phys. Rev. Lett.\/} {\bf 81}, 4140--4143.

\bibitem[Willis(2017)]{openpipeflow}
{\sc Willis, A.~P.} 2017 {Openpipeflow}: {Pipe} flow code for incompressible
  flow. {\em Tech. Rep.\/}. U. Sheffield, {\tt {Openpipeflow.org}}.

\bibitem[Willis \& Kerswell(2009)]{WillKer09}
{\sc Willis, A.~P. \& Kerswell, R.~R.} 2009 Turbulent dynamics of pipe flow
  captured in a reduced model: puff relaminarisation and localised edge states.
  {\em J. Fluid Mec.\/} {\bf 619}, 213--233, \arXiv{0712.2739}.

\bibitem[Willis {\em et~al.\/}(2016)Willis, Short \& Cvitanovi{\'c}]{WiShCv15}
{\sc Willis, A.~P., Short, K.~Y. \& Cvitanovi{\'c}, P.} 2016 Symmetry reduction
  in high dimensions, illustrated in a turbulent pipe. {\em Phys. Rev. E\/}
  {\bf 93}, 022204, \arXiv{1504.05825}.

\bibitem[Zammert \& Eckhardt(2015)]{ZamEck15}
{\sc Zammert, S. \& Eckhardt, B.} 2015 Crisis bifurcations in plane
  {Poiseuille} flow. {\em Phys. Rev. E\/} {\bf 91}, 041003.

\end{thebibliography}

    \ifboyscout
	    \newpage
	    \input flotsam
    \fi %end of internal draft switch

\end{document}